\documentclass[12pt]{article}
\usepackage{graphics}
\begin{document}

\newcommand{\hs}{\hspace{1cm}}
\newcommand{\be}[1]{\begin{equation}\label{#1}}
\newcommand{\BE}{\begin{equation}}
\newcommand{\ee}{\end{equation}}
\newcommand{\bee}{\begin{eqnarray}}
\newcommand{\eee}{\end{eqnarray}}

\newcommand{\uu}[1]{\underline{\underline{#1}}}

\newcommand{\RP}{R_P}
\newcommand{\RN}{R_N}

\title{Comb-like polymers inside nanoscale pores}
\author{C. Gay~(1), E. Rapha\"el~(2)\\ 
(1) Laboratoire CNRS - Atofina (UMR 167),\\
95, rue Danton, B.P. 108, 
92303 Levallois-Perret cedex, France,\\
cgay@pobox.com\\ 
(2) Coll\`ege de France, Physique de la Mati\`ere Condens\'ee\\
(URA 792 du CNRS),\\
11, place Marcelin Berthelot, 
75231 Paris cedex 05, France,\\
elie.raphael@college-de-france.fr}
\date{}
\maketitle

\section{Introduction}
A new method of polymer characterization, based on permeation studies
using nanoscale pores, was recently proposed by
Brochard and de Gennes \cite{PGGFB}.
In the present paper, we study how this
method, initially developped for star polymers,
can be extended to comb-like polymers \cite{PGG3}.
The present study is based on the Flory free energy approach 
and therefore does not provide the more complete description 
of chain conformations that can be obtained in simple geometries 
through self-consistent field methods. 
Furthermore, our scaling analysis leaves numerical factors 
undetermined. 
The crossovers between the different regimes we obtain 
are expected to extend over a factor of two or so in the parameters.

\section{A comb in a good solvent : from worm to star}

Consider a comb homopolymer made of $n$ segments, each containing
$N$ monomers along the backbone and $P$ monomers in a side-chain.
The various possible chain conformations in good solvent can be
derived from a classical Flory free energy approach. They are
described by the quantities listed in the following
Table and are pictured
on the Figure.
$R$ is the radius of gyration of the whole molecule, 
$R_{nN}$ is that of the main chain 
and $R_P$ is that of the side-chains.
Where it is relevant, 
$R_C$, $n_C$ and $P_C$ are the size, number of side-chains 
or number of monomers from each side-chain 
in the central region of the molecule 
described later in the text. \newline

\begin{tabular}{|c|c|c|c|}
 \hline
  \begin{tabular}{c}
    Overall \\
    molecule \\
    conformation \\
  \end{tabular} & \begin{tabular}{c}
    Radius \\
    of \\
    gyration \\
  \end{tabular}
  & \begin{tabular}{c}
    Backbone \\
    conformation \\
    REGIME \\
  \end{tabular} & \begin{tabular}{c}
    Other \\
    characteristics \\
  \end{tabular}
  \\ \hline
 \begin{tabular}{c}
    Decorated \\
    Chain $R=R_{nN}$\\
  \end{tabular} &
  \begin{tabular}{c}
    $R=a(nN)^{3/5}$ \\
    (Flory) \\
  \end{tabular} & \begin{tabular}{c}
    flexible \\
    DC \\
  \end{tabular} & $R_P=aP^{3/5}$ \\ \hline
  \begin{tabular}{c}
    Worm $R=R_{nN}$\\
    $R_C=R_P$ \\
    $R=R_C\left(\frac{n}{n_C}\right)^{3/5}$ \\
  \end{tabular}
  & \begin{tabular}{c}
    $R=an^{3/5}P^{2/5}N^{1/5}$ \\
  \end{tabular}
 &  \begin{tabular}{c}
      flexible \\
      FBW \\
    \end{tabular}
    & \begin{tabular}{c}
    $P=n_C^2N$\\\hline
    $R_P=aP^{3/5}n_C^{1/5}$ \\
    $=aP^{7/10}N^{-1/10}$ \\
  \end{tabular}
 \\ \hline
  id. & $R=an^{3/5}N^{1/2}P^{3/10}$
  & \begin{tabular}{c}
      stretched \\
      SBW \\
    \end{tabular} &
    \begin{tabular}{c}
    $P=n_C^{4/3}N^{5/3}$\\\hline
    $R_P=an_CN$\\
    $=aP^{3/5}n_C^{1/5}$ \\
    $=aP^{3/4}N^{-1/4}$ \\
  \end{tabular}
 \\ \hline
  \begin{tabular}{c}
    Star $R=R_P$\\
    $n_C=n$ \\
    $R_C=R_{nN}$ \\
  \end{tabular}
& \begin{tabular}{c}
    $R=aP^{3/5}n^{1/5}$ \\
    (Daoud-Cotton) \\
  \end{tabular}
 &     \begin{tabular}{c}
      flexible \\
      FBS \\
    \end{tabular}
    & \begin{tabular}{c}
    $P_C=n^2N$\\\hline
    $R_C=aP_C^{3/5}n^{1/5}$ \\
    $=aN^{3/5}n^{7/5}$ \\
  \end{tabular} \\ \hline
  id. & id. &   \begin{tabular}{c}
      stretched \\
      SBS \\
    \end{tabular} &
    \begin{tabular}{c}
    $P_C=n^{4/3}N^{5/3}$\\\hline
    $R_C=anN$\\
    $=aP_C^{3/5}n^{1/5}$ \\
  \end{tabular} \\ \hline
\end{tabular}\newline\newline
These various regimes can be understood quite simply.
When side-chains are short ($P<N$), the whole comb behaves like a
simple chain of $nN$ monomers, decorated with swollen side-chains.
The gyration radii of both the side-chains and the overall
molecule follow a Flory scaling law (regime DC).
\begin{figure}
\centering
\resizebox{10cm}{!}  
{\includegraphics*[3.5cm,9.5cm][17cm,19cm]{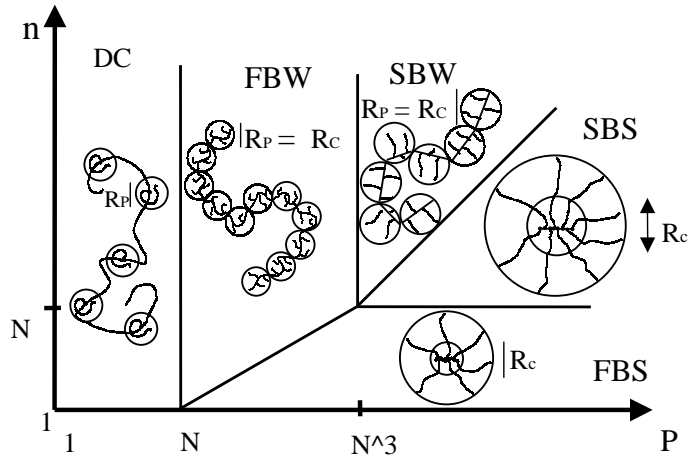}}  
\caption{Behavior of a comb-like polymer (made of $n$ segments, each containing
$N$ monomers along the backbone and $P$ monomers in a side-chain) in
a good solvent.}
\end{figure}
If side-chains are very long, they expand far away from the
backbone, and the overall conformation of the comb molecule is
that of a star with $n$ arms of length $P$ (regimes FBS and SBS).
Its radius of gyration is that given by Daoud and
Cotton~\cite{DAOUDCOTTON}. The central region of the star contains
the backbone (core of size $R_C$). If the number of monomers along
the backbone between side-chains is small ($N<n$), the backbone is
completely stretched ($R_C=anN$). This is the stretched backbone
star regime (SBS). For larger values of $N$, the backbone is
weakly stretched (flexible backbone star (FBS) regime). The
smaller core size $R_C$ can be obtained from the following Flory
free energy which takes into account the side-chain and backbone
elastic free energy and the excluded volume interactions in the
core region:
\begin{equation}\label{RCPC}
\frac{F}{kT}=\frac{R_C^2}{nNa^2} +n\frac{R_C^2}{P_Ca^2}
+\frac{(nP_C)^2a^3}{R_C^3}
\end{equation}
where $P_C\leq P$ is the number of monomers from each side-chain
that are within the core. Note $P_C$ can also be obtained from the
continuity of the osmotic pressure $kT/\xi^3$ between the core
interior (with $\xi=a\phi^{-3/4}$ where $\phi\simeq
nP_Ca^3/R_C^3$) and the core boundary, where the arms of the star
depart from one another (where $R_C^2=n\xi^2$, see
reference~\cite{DAOUDCOTTON}).


If the side-chain length $P$ is reduced, then at some point, the
side-chains do not extend beyond the core region, precisely, when
$P=P_C$. The star is now reduced to its core. Let $n_C=n$ be the
number of segments. If the total number $n$ of side-chains is now
increased beyond $n_C$, the molecule develops into a self-avoiding
walk of such cores whose size is $R_C$ (strongly or weakly
stretched) and thus resembles a wiggling worm 
(regime~SBW or~FBW) 
whose radius of gyration is therefore equal to
$R=R_C\,(n/n_C)^{3/5}$ 
(where $n_C$ is the number of segments per core). 
This result is confirmed by the fact 
that in the FBW regime (flexible backbone), 
the overall radius of gyration
$R=R_C\,(n/n_C)^{3/5}=an^{3/5}N^{1/5}P^{2/5}$ 
can be obtained directly by minimizing
a simple Flory free energy~\cite{NOTE}:
\begin{equation}
\frac{F}{kT}=\frac{R^2}{nNa^2}+\frac{(nP)^2a^3}{R^3}
\end{equation}

\section{Confined comb and injection threshold}

As shown above, depending on the molecular parameters, the overall
conformation of a comb in good solvent is similar to that of a
swollen linear chain if side-chains are short or intermediate in
length, and it is similar to that of a star if side-chains are
long. The way in which such a comb may permeate through a nanopore
(diameter $D$) due to a solvent flow is therefore expected to be
quite different in these various conformation regimes. Indeed, the
critical solvent flow that is necessary to force a linear chain
into the nanopore is molecular weight independent~\cite{PGGLIN}~:
\begin{equation}
J_c=\frac{kT}{\eta}\equiv J_{c1}\label{JcLIN}
\end{equation}
This applies to the comb molecule in the decorated chain regime
($P<N$) or in the weak confinement worm regime ($R_C<D<R$).

Conversely, the threshold current for stars is sensitive to the
number of arms and to their molecular weight, depending on the
nanopore diameter ~\cite{PGGFB,PGG3}~:
\begin{eqnarray}
J_c^{star}&=&J_{c1}\hs{\rm if}\hs D<aN/n^{3/2}\label{JcSTAR0}\\
J_c^{star}&=&J_{c1}\cdot n\left(\frac{D}{Na}\right)^{2/3}\hs{\rm if}\hs
aN/n^{3/2}<D<aN/n^{3/2}\label{JcSTAR1}
\end{eqnarray}
where in the first case the molecule penetrates the pore with just
one forward arm at first, 
and in the second case the number of
forward arms is $n_0\simeq n^{1/2}(D/Na)^{1/3}$. 
These results
apply as such to the comb molecule in the star regimes (FBS and
SBS) as long as the core of the comb is smaller than the tube
diameter ($R_C<D<R$), and even somewhat beyond $D\simeq R_C$, as
we shall see.

We now wish to investigate situations 
where the pore diameter is smaller, 
in order to determine to what extent comb molecules 
can be characterized by such a method.
The first step is to determine the conformation of a comb molecule
in such intermediate regimes, once it has completely entered the
nanopore. As can be seen from Figure~2, if side-chains are long,
the total length $L$ of the object is essentially their length
$L_P$ which can be deduced from a Flory argument. The elastic
energy of the side chains is balanced by the excluded volume free
energy~:
\begin{equation}
\frac{F}{kT}\simeq
n\frac{L_P^2}{P\,a^2}+\frac{(nP)^2\,a^3}{L_P\,D^2}
\end{equation}
Minimizing this energy yields the length of the comb molecule
which is proportional to the side-chain molecular weight~$P$~:
\begin{equation}\label{LCONFSTAR}
L\simeq L_P\simeq a\,P\,n^{1/3}\,(a/D)^{2/3}
\end{equation}
Conversely, if the side-chains are shorter, the length of the
object is the length of the backbone and its elasticity is the
dominant elastic contribution~:
\begin{equation}
\frac{F}{kT}\simeq
\frac{L_{nN}^2}{nN\,a^2}+\frac{(nP)^2\,a^3}{L_{nN}\,D^2}
\end{equation}
The length of the comb molecule is now proportional to the number
$n$ of molecule segments~:
\begin{equation}\label{LCONFWORMF}
L\simeq L_{nN}\simeq a\,n\,P^{2/3}\,N^{1/3}\,(a/D)^{2/3}
\end{equation}
The crossover between equations~(\ref{LCONFSTAR})
and~(\ref{LCONFWORMF}) occurs for $P\simeq N\,n^2$ and provides
the basic ingredient for a finer description of the comb
conformation, as we now see.

Consider a star in the regime of equation~(\ref{LCONFSTAR}) and
suppose that side-chains $P$ are now shorter. The central region
of the molecule is not affected. In fact, it will start to be
affected when the side chain length has decreased to the backbone
length $L_{nN}$, i.e., when we reach the worm regime of
equation~(\ref{LCONFWORMF}). In other words, the length $L_{nN}$
of the central region is given by equation~(\ref{LCONFSTAR}) where
$P$ is chosen as $N\,n^2$, i.e. $L_{nN}\simeq
a\,N\,n^{7/3}\,(a/D)^{2/3}$.

\begin{figure}
\centering
\resizebox{\textwidth}{!}  
{\includegraphics*[0.5cm,10cm][19.5cm,24.3cm]{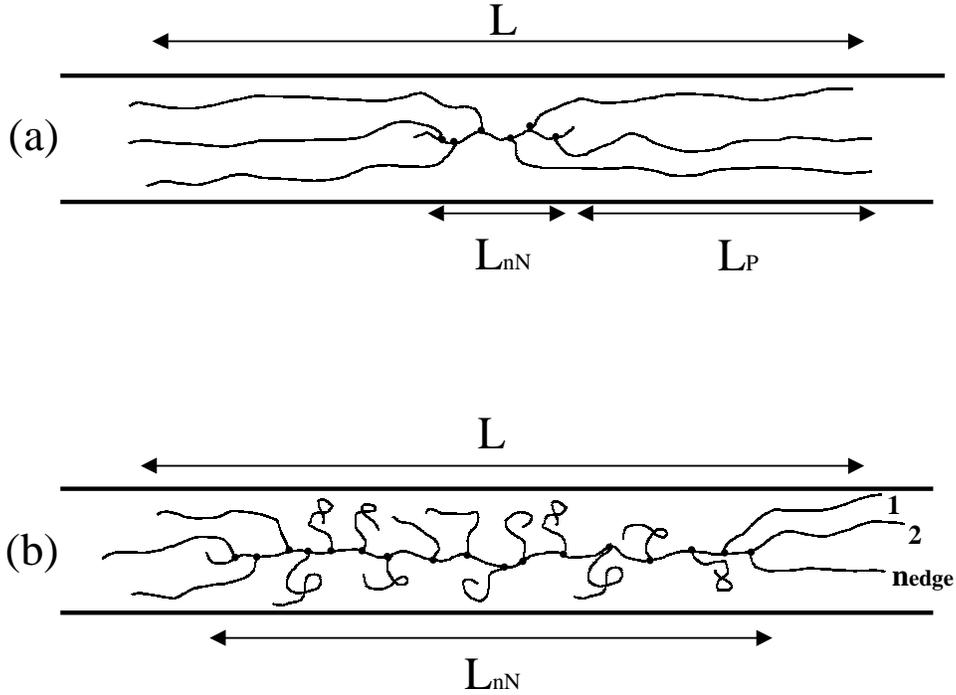}}  
\caption{ Conformation of a comb-like polymer molecule inside a nanopore. 
(a) Star-like conformation~: the overall length $L$ 
is comparable to the arm length $L_P$,
and the backbone (length $L_{nN}\ll L$) 
may be weakly or strongly stretched 
(regime C-FBS or C-SBS). 
(b) Worm conformation~: 
the overall length of the molecule 
is essentially that of the backbone ($L\simeq L_{nN}$) 
which is either weakly or strongly stretched 
(regime C-FBW or C-SBW), 
and about $n_{edge}\simeq n_C$ arms 
stretch away from the backbone 
along the tube axis (edge effect).}
\end{figure}
Let us now consider a worm in the regime of
equation~(\ref{LCONFWORMF}). The side-chains located near the end
of the backbone have extra space and are extended outwards. How
far into the molecule does this edge effect penetrate can be
estimated by decreasing the number $n$ of molecule segments~: when
$n$ is such that $P\simeq N\,n^2$, the molecule crossovers to the
star regime and the edge effect has invaded the whole molecule
conformation. Turning back to the worm regime, this yields the extension
$L_P\simeq a\,P\,n_{edge}^{1/3}\,(a/D)^{2/3} \simeq
a\,P^{7/6}\,N^{-1/6}\,(a/D)^{2/3}$ of the edge effect, where the
number of side-chains involved is given by~:
\begin{equation}\label{NEDGEWORMF}
n_{edge}\simeq (P/N)^{1/2}
\end{equation}

If the worm backbone is stretched ($L=L_{nN}\simeq anN$), 
the same approach yields the detailed conformation~:
\begin{equation}
n_{edge}\simeq(P/N)^{3/2}a/D\hs
L_P\simeq P^{3/2}/N^{1/2}\,a^2/D
\end{equation}

It can be seen from all these results 
that in the confined geometry, 
the extension of the various regimes are altered 
because confinement may stretch the backbone completely~:
the limit between 
the confined flexible backbone worm regime (C-FBW) 
and the stretched backbone worm (C-SBW) 
is now $P\simeq N\,D/a$ instead of $P\simeq N^3$, 
the limit between the flexible backbone star (C-FBS) 
and the stretched backbone star (C-SBS) 
is now $n\simeq (D/a)^{1/2}$ instead of $n\simeq N$, 
and the limit between the stretched backbone 
star and worm (C-SBS and C-SBW) 
is $n=(P/N)^{3/2}\,a/D$ instead of $n\simeq P^{3/4}/N^{5/4}$. 

The results of reference~\cite{PGGFB} 
for the threshold current (see equations~\ref{JcSTAR0}, 
\ref{JcSTAR1}) 
can be easily transposed here. 
Indeed, the edge of the confined worm 
has the same conformation as one half of a star 
with $2n_{edge}$ arms. 
This is also true before the molecule 
enters the pore since the expression for $n_{edge}$ 
is the same as that for $n_C$. 
The threshold current for the worm 
can is thus obtained simply by using 
the equivalent number of star arms 
$n_{edge}$ (flexible worm regime C-FBW, 
equation~\ref{NEDGEWORMF}) 
instead of $n$ in equations 
(\ref{JcSTAR0},\ref{JcSTAR1})~:
\begin{eqnarray}
J_c^{C-FBW}&=&J_{c1}\hs{\rm if}\hs \frac{D}{a}<\frac{N^{7/4}}{P^{3/4}}\\
J_c^{C-FBW}&=&J_{c1}\cdot
\frac{P^{1/2}}{N^{1/2}}\left(\frac{D}{Na}\right)^{2/3}\hs{\rm
if}\hs \frac{D}{a}>\frac{N^{7/4}}{P^{3/4}}\label{JcWORMF1}
\end{eqnarray}
Similarly, for the stretched backbone case 
(regime C-SBW)~:
\begin{eqnarray}
J_c^{C-SBW}&=&J_{c1}\hs{\rm if}\hs \frac{D}{a}>\frac{P^{9/2}}{N^{13/2}}\\
J_c^{C-SBW}&=&J_{c1}\cdot
\frac{P^{3/2}}{N^{5/2}}\left(\frac{Na}{D}\right)^{1/3}\hs{\rm
if}\hs \frac{D}{a}<\frac{P^{9/2}}{N^{13/2}}\label{JcWORMS1}
\end{eqnarray}

By comparing the various threshold current values obtained 
with Figure~1, we find the regimes 
where they are relevant.
They essentially correspond to the 
confined regimes described above.
Indeed, as soon as $R>D>N\,a$, 
the star threshold of equation~(\ref{JcSTAR1}) 
applies to regimes C-FBS and C-SBS, and the worm thresholds 
of equations~\ref{JcWORMF1} and~\ref{JcWORMS1} 
apply in regimes C-FBW and C-SBW respectively.
If the pore diameter is smaller ($aN^{4/7}<D<Na$), 
the non-sensitive regime $J_c=J_{c1}$
includes not only the decorated chain regime DC, 
but also a portion of the star and worm regimes 
(namely $n<(Na/D)^{2/3}$ 
and $P<N^{7/3}(a/D)^{4/3}$).
If the diameter is even smaller ($D<aN^{4/7}$), 
the non-sensitive regime $J_c=J_{c1}$ 
includes $n<(Na/D)^{2/3}$ 
and $P<N^{13/9}(D/a)^{2/9}$,
and equations~(\ref{JcSTAR1},\ref{JcWORMS1}) 
are valid beyond these limits.\\

\noindent {\Large {\bf Acknowledgements}}\\
We thank Professor P.-G. de Gennes for very helpful discussions and comments\\
\\





\begin{thebibliography}{10}
\bibitem{NOTE}
The fact that we obtain $R=R_C\,(n/n_C)^{3/5}$ 
through the overall Flory argument for the whole molecule 
implies that the persistence length $\lambda$ 
of the worm is of the order of its radius $R_C$. 
This question is by no means obvious 
and has been much debated in recent years. 
It is addressed more fully in a separate study~\cite{PERSIST}, 
whose essential argument can be expressed qualitatively 
in the following terms.
One could imagine that the osmotic pressure
of the side-chains could induce a higher worm rigidity~: 
consider that the worm of radius $R_C$ behaves like 
some continuous elastic medium whose modulus is given by the
osmotic pressure $\pi$, then the bending rigidity $\mu$ is on the
order of $\pi\,R_C^4$. The persistence length is then
given by $\lambda\simeq\mu/kT\simeq R_C^4/\xi^3$, 
which can be much larger than $R_C$. 
Such an elastic worm model is in fact too crude, 
since side-chains are 
free to rotate about the main chain when the worm is subjected to
flexion, in order to relieve compression on the inner side and
expansion on the outer side. 
Similarly, the chains are free to adjust 
how far they extend away from the backbone. 
We show in reference~\cite{PERSIST} 
that the combination of these two effects 
lead to negligible bending rigidity 
on length scales larger than $R_C$.
\end{thebibliography}
\end{document}